\documentclass[twocolumn,secnumarabic,amssymb, nobibnotes, superscriptaddress, nofootinbib,aps,prd]{revtex4-2}
\usepackage[colorlinks,citecolor=blue,linkcolor=blue,anchorcolor=blue,filecolor=blue, urlcolor=blue]{hyperref}
\usepackage{natbib}

\usepackage{amsmath}
\usepackage{adjustbox}
\usepackage{graphicx}
\usepackage{booktabs}       
\usepackage{amsfonts}       
\usepackage{nicefrac}       
\usepackage{microtype}
\usepackage{cleveref}
\usepackage{xcolor}
\usepackage{multirow}
\usepackage{makecell}
\usepackage{comment}
\usepackage{todonotes}
\usepackage{array}
\usepackage{orcidlink}
\usepackage{float}
\usepackage[normalem]{ulem}

\begin{document}

\title{Proto-neutron Stars with Dark Matter Admixture: A Single-Fluid Approach}

\author{Adamu Issifu~\orcidlink{0000-0002-2843-835X}} 
\email{ai@academico.ufpb.br}
\affiliation{Departamento de F\'isica, Instituto Tecnol\'ogico de Aeron\'autica, DCTA, 12228-900, S\~ao Jos\'e dos Campos, SP, Brazil} 
\affiliation{Laborat\'orio de Computa\c c\~ao Cient\'ifica Avan\c cada e Modelamento (Lab-CCAM)}

\author{D\'ebora P. Menezes~\orcidlink{0000-0003-0730-6689}}
\email{debora.p.m@ufsc.br}
\affiliation{Departamento de F\'isica, CFM - Universidade Federal de Santa Catarina; \\ C.P. 476, CEP 88.040-900, Florian\'opolis, SC, Brazil.}

\author{Tobias Frederico~\orcidlink{0000-0002-5497-5490}} 
\email{tobias@ita.br}
\affiliation{Departamento de F\'isica e Laborat\'orio de Computa\c c\~ao Cient\'ifica Avan\c cada e Modelamento (Lab-CCAM), Instituto Tecnol\'ogico de Aeron\'autica, DCTA, 12228-900, S\~ao Jos\'e dos Campos, SP, Brazil}

\date{\today}
\begin{abstract}
This work investigates the impact of dark matter (DM) on the microscopic and macroscopic properties of proto-neutron stars (PNSs). We employ a single-fluid framework in which DM interacts with ordinary matter (OM) via the Higgs portal and remains in thermal equilibrium through non-gravitational interactions. Using a quasi-static approximation, we analyze the evolution of PNSs during the Kelvin–Helmholtz phase by varying the DM mass while keeping the entropy per baryon and lepton fraction fixed. Our results show that DM absorbs thermal energy from the stellar medium without efficient re-emission, thereby altering neutrino emission and affecting the star’s thermal evolution history. Furthermore, neutrinos contribute significantly to pressure support in the PNS phase, inhibiting DM mass accretion during neutrino-trapped stages. Based on the requirement to satisfy the observed $2\,\rm M_\odot$ neutron star mass constraint and to maintain consistency with supernova remnant data, we suggest an upper limit of $m_\chi \leq 0.62\,\rm GeV$ for the DM mass that can accrete in evolving PNSs, within the model framework. In contrast, we established that cold neutron stars (NSs) can support higher DM masses without compromising equilibrium stability, owing to increased central density, enhanced gravitational binding energy, and reduced thermal pressure.
\end{abstract}

\maketitle

\section{Introduction}
 
Evidence from galactic rotation curves, large-scale cosmological structures, gravitational lensing, and other observations strongly suggests the presence of DM in the universe. Current cosmological models estimate that the universe is composed of approximately 26.4\% DM and 67.6\% dark energy, together accounting for about 94\% of total mass-energy content in the universe~\cite{Planck:2018vyg, Massey:2010hh, Lelli:2016zqa}. Despite decades of efforts to directly detect DM in terrestrial laboratories and colliders, no results have been obtained so far~\cite{PandaX-II:2017hlx, CRESST:2019jnq, XENON:2018voc}. As a result, the true nature of dark matter, its interactions, coupling strength, and mass remain speculative. Given the strong evidence for its existence, several theoretically motivated models treat DM as a fundamental particle.  However, the search for it remains an area of intense research, with decades of direct and indirect detection experiments conducted in laboratories, colliders, and observatories~\cite{Klasen:2015uma, Schumann:2019eaa, Liu:2017drf}.

The complexity of determining the nature and properties of DM in the universe makes its modeling difficult. Therefore, there are two main approaches for theoretically investigating the effects of DM on compact objects such as NSs. It can be studied using the two-fluid Tolman-Oppenheimer-Volkoff (TOV) formalism, where DM and OM are assumed to interact only gravitationally~\cite{Leung:2011zz, Kain:2021hpk, Collier:2022cpr, Shakeri:2022dwg, Miao:2022rqj, Rutherford:2022xeb, Das:2020ecp}. This formalism follows the conventional assumption that DM interacts weakly with OM. Consequently, the equations of state (EoS) for DM and OM are calculated separately, with their interaction introduced gravitationally through the two-fluid TOV framework. The second approach, on the other hand, is the single-fluid formalism, where non-gravitational interactions between DM and OM are assumed. This approach has several advantages, including simplicity and computational efficiency. It leads to faster simulations and reduced computational costs, as it makes the TOV equations easier to solve numerically without considering two interpenetrating fluids~\cite{Lenzi:2022ypb, Lourenco:2022fmf, Das:2018frc, Das:2020vng, Lopes:2024ixl}. 

The existence of multiple DM species has been theorized, and several experimental efforts are currently underway to investigate their properties. Among the leading theoretical candidates are weakly interacting massive particles (WIMPs)~\cite{Arcadi:2017kky, LUX:2015abn}, feebly interacting massive particles (FIMPs)~\cite{Antel:2023hkf, Bernal:2017kxu}, neutralinos~\cite{Gelmini:2006pw, Boehm:2013qva}, and axions~\cite{Chadha-Day:2021szb, OHare:2024nmr}, among others. WIMPs are particularly compelling, as they are believed to have existed in thermal equilibrium in the early universe, undergoing frequent pair annihilations. As the universe expanded and cooled, the annihilation rate dropped below the cosmic expansion rate, causing WIMPs to decouple, a process known as freeze-out, ultimately resulting in the thermal relic DM abundance~\cite{vonHarling:2014kha}. The appeal of WIMPs lies in the fact that their relic abundance can be directly linked to the weak interaction scale, making them promising candidates for direct detection in terrestrial experiments.
The challenges in detecting DM through direct detection have led to an increased focus on indirect searches through astrophysical observations. One such avenue is the study of DM accretion in NSs, where their dense cores can efficiently trap DM particles~\cite{Bramante:2023djs, Bramante:2017xlb, Thakur:2023aqm}. The rate at which NS accumulates DM and the effects on its microscopic and macroscopic structures depend on the type of DM (bosonic or fermionic) and its properties. For example, the accretion of self-annihilating DM~\cite{Jungman:1995df, Fermi-LAT:2015att}  can heat the star, altering its cooling curves compared to conventional models. On the other hand, non-self-annihilating DM, such as asymmetric DM~\cite{Petraki:2013wwa, Zurek:2013wia} trapped in the star, can modify its mass-radius relation, oscillatory properties, tidal deformability, and gravitational wave signature~\cite{Nelson:2018xtr, Bramante:2013hn}.

In recent years, significant advances have been made in NS observations. In particular, the detection of the gravitational wave event GW170817 from the merger of binary NSs~\cite{LIGOScientific:2017vwq}, along with data from the NICER X-ray observatory, has placed stringent constraints on the EoS of NS matter. The simultaneous measurement of the mass and radius of pulsars PSR J0030$+$0451~\cite{riley2019, Miller:2019cac} and PSR J0740$+$6620~\cite{riley2021, Miller:2021qha}, along with constraints from GW170817, provides valuable information on the maximum mass and radius that an NS can support before collapsing. Therefore, a realistic modern EoS must satisfy these observational constraints. 

This work employs a fermionic dark matter candidate that interacts with nucleons through the Higgs portal mechanism~\cite{Bell:2019pyc, Arcadi:2019lka}. The choice of this model remains phenomenologically viable within certain parameter spaces, as proposed by direct detection experiments, collider searches, and nuclear physics considerations (see a review in~\cite{Misiaszek:2023sxe}). We study the evolution of NSs from their birth as neutrino-rich objects admixed with DM to their maturity as cold-catalysed neutrino-poor DM-admixed NS (DANS), ensuring that the constraints from direct detection experiments and collider searches are satisfied by the model. The OM component is described using a relativistic NS model with a density-dependent coupling, which is adjusted according to the DDME2 parameterisation. We determine the EoS along with the temperature profiles and particle distributions by fixing the entropy per baryon ($s_B$) and the lepton fraction~($Y_{L,e}$), using the thermodynamic conditions relevant for PNS evolution~\cite{Prakash:1996xs, Janka:2006fh}. In the fixed $s_B$ regime, we evaluated the integrals over the entire momentum space and varied the DM mass to determine its effect on the microscopic and macroscopic properties of the star. The strength of the Higgs-nucleon coupling is investigated, and the variations in the DM density and the DM chemical potential with the baryon density in the visible sector are also studied. 

The cooling effect of DM admixed NSs has been widely investigated through cooling curves using a thermal balance equation with a redshifted surface temperature~\cite{Bhat:2019tnz, Avila:2023rzj}. DM thermalization in NSs has also been studied in~\cite{Bertoni:2013bsa, Bell:2023ysh}. So far, no work has been found in the literature, to the best of our knowledge, that investigates the evolution of PNSs considering thermal equilibrium between DM and OM, which consistently derives the EoS for DANSs from the neutrino-rich phase to the neutrino-poor regime in the single-fluid formalism, as we propose in this work. We should mention that Ref.~\cite{Issifu:2024htq} explores the microscopic and macroscopic properties of PNSs admixed with DM within a two-fluid framework. For the first time, that study demonstrates how gravitational interactions shape particle distribution and temperature profiles in PNSs using a fermionic mirrored DM model. However, it is necessary to investigate how the interaction between DM and OM can change this scenario. Consequently, the present work aims to provide a basis for comparing how different DM-OM interaction types influence stellar properties.


The paper is structured as follows: Sec.~\ref{mph} introduces the EoSs for OM and DM, where DM interacts non-gravitationally with OM via the Higgs portal, and discusses the relevant microphysics. This section is further divided into Subsecs.~\ref{om} and \ref{dm}, which provide detailed discussions of the OM and DM models, respectively. Sec.~\ref{r} presents our results and analysis, while Sec.~\ref{c} concludes the paper.

\section{Microphysics}\label{mph}

\subsection{Hadronic Matter}\label{om}
The EoS of NS matter is governed by the field-theory-motivated relativistic mean-field (RMF) approximation, where the nucleon-nucleon interaction is modeled using massive mesons, $i = \sigma, \omega, \rho$. Here, $\sigma$ is a scalar meson, $\omega$ is a vector-isoscalar meson, and $\rho$ is a vector-isovector meson. The Lagrangian density of the model is given by~\cite{Menezes:2021jmw}:

\begin{align}
\mathcal{L}_{\rm OM}{}&= \mathcal{L}_{\rm H} + \mathcal{L}_{\rm m}+ \mathcal{L}_{\rm L},\\
 \mathcal{L}_{\rm H}{}&=  \sum_{N=p,n} \bar \psi_N \Big[  i \gamma^\mu\partial_\mu - \gamma^0  \big(g_{\omega N} \omega_0  +  g_{\rho N} I_{3} \rho_{03}  \big)\nonumber\\
 &- \Big( m_N- g_{\sigma N} \sigma_0 \Big)  \Big] \psi_N, \label{h}\\
\mathcal{L}_{\rm m}&= - \frac{1}{2} m_\sigma^2 \sigma_0^2  +\frac{1}{2} m_\omega^2 \omega_0^2   +\frac{1}{2} m_\rho^2 \rho_{03}^2,\label{m}\\
 \mathcal{L}_{\rm L}& = \sum_L\Bar{\psi}_L\left(i\gamma^\mu\partial_\mu-m_L\right)\psi_L\label{l},
\end{align}
where $\psi_N$ is the Dirac-type field for the nucleons, $g_{iN}$ is the meson-nucleon coupling, $m_N=938$ MeV is the nucleon mass, $m_{i}$ is the meson mass, and $I_3=\pm 1/2$ is the isospin projection. Also, $\psi_L$ is the lepton field with the index $L$ running over all the leptons present in the matter, and $m_L$ is the lepton mass. $\mathcal{L}_{\rm OM}$ is the Lagrangian density of the OM sector comprising hadrons $\mathcal{L}_{\rm H}$, mesons $\mathcal{L}_{\rm m}$ and leptons $\mathcal{L}_{\rm L}$. The strength of the meson-nucleon coupling and the meson masses are based on the DDME2 parameterization and have been presented in Table~\ref {T}. Since NSs are physically observable objects, we ensure that the stellar matter is chemically stable and charge-neutral. So the leptons are introduced as free Fermi-gas to balance the proton ($p$) charge. The couplings are adjusted by~\cite{Roca-Maza:2011alv}:
\begin{align}
    g_{i b} (n_B) &= g_{ib} (n_0)a_i  \frac{1+b_i (\eta + d_i)^2}{1 +c_i (\eta + d_i)^2}, \label{cp}\\
    g_{\rho b} (n_B) &= g_{\rho b} (n_0) \exp\left[ - a_\rho \big( \eta -1 \big) \right],
\end{align}
here, $n_B$ is the total baryon density and $\eta = n_B/n_0$, where $n_0 = 0.152\rm fm^{-3}$ is the nuclear saturation density under this framework. The model parameters $a_i, b_i, c_i$ and $d_i$ were fitted to experimental bulk nuclear properties, and their corresponding values are presented in Table~\ref {T}. Other nuclear properties determined under this model framework are: $E_B = -16.14$~MeV~(binding energy), $K_0 = 251.9$~MeV~(incompressibility), $J = 32.3$~MeV~(symmetry energy), and $L_0 =51.3$~MeV~(symmetry energy slope). These properties agree well with symmetric nuclear properties determined by other authors~\cite{Reed:2021nqk, Lattimer:2023rpe} and references therein.

For PNS matter, we consider charge-neutral, $\beta$-equilibrated neutrino-trapped, and neutrino-transparent matter at different stages of the star's evolution. In neutrino-trapped matter, electrons ($e$) and their corresponding electron neutrinos ($\nu_e$) are considered. On the other hand, muons become relevant only after all neutrinos have escaped from the stellar core, while tau leptons are considered too heavy to be present, as dictated by supernova physics~\cite{Prakash:1996xs}. An isentropic EoS using the RMF approximation with density-dependent couplings is then determined by calculating the energy-momentum tensor components 
from $\mathcal{L}_{\rm OM}$. A detailed derivation of the EoS can be found in~\cite{Issifu:2023qyi, Malfatti:2019tpg, Raduta:2020fdn, Issifu:2024fuw, Marques:2017zju} and references therein.

\subsection{Dark Matter}\label{dm}
For the DM component, we consider neutralino as a fermionic DM candidate that interacts with nucleons via the Standard Model (SM) Higgs portal for a comprehensive study of DM admix NSs (DANSs). The Lagrangian density governing this interaction is given by~\cite{Lourenco:2022fmf, Das:2021hnk, Lenzi:2022ypb, Panotopoulos:2017idn, Lopes:2023uxi}: 
\begin{align}
\mathcal{L}_{\rm DM} &= \bar{\chi}(i \gamma^\mu \partial_\mu - (m_\chi -g_h h))\chi \nonumber\\
&+ \frac{1}{2}(\partial^\mu h \partial_\mu h - m_h^2 h^2) + \sum_{N}\dfrac{fm_N}{v} \bar{\psi}_Nh\psi_N, \label{FDMEOS}
\end{align}
with $\chi$ and $\psi_N$ representing the DM and the nucleon fields, respectively. $h$, $m_\chi = 200$ GeV, $m_h = 125$ GeV, and $g_h$ denote the Higgs field, neutralino mass, Higgs mass, and DM-Higgs coupling strength, respectively; $f$ is the Higgs-nucleon form factor, and $v = 246$ GeV is the Higgs vacuum expectation value. We fix $f=0.35$ based on the estimated value from lattice QCD-motivated results~\cite{Czarnecki:2010gb}, which aligns with the MILC results~\cite{Toussaint:2009pz} reported in~\cite{Djouadi:2011aa} and also falls within the range $0.26-0.63$ determined by ATLAS~\cite{ATLAS:2015ciy}. Meanwhile, $g_h = 0.07$ has been determined as a reasonable choice~\cite{Das:2021hnk} within the permitted range of values $0.001\leq g_h \leq 0.1$~\cite{Panotopoulos:2017idn}. Using these model parameters the spin-independent scattering cross-section of the nucleon with DM has been determined to be $9.70\times 10^{-46} \rm cm^2$~\cite{Das:2021hnk} which is consistent with the constraints imposed by the null experiment results: XENON-1T~\cite{XENON:2015gkh}, PandaX-II~\cite{PandaX-II:2016vec}, PandaX-4T~\cite{PandaX-4T:2021bab} and LUX~\cite{LUX:2016ggv} within 90\% confidence level. This result also falls within the range of the WIMP-nucleon cross-section proposed at the LHC, which is between $10^{-40}$ to $10^{-50}$\,cm$^2$~\cite{Djouadi:2011aa}. 

In the zero-temperature regime, when the star is cold and catalyzed, the Fermi momentum ($k_f^D$) of the DM is varied between 20 MeV to 60 MeV, regulating the amount of DM that the star can accumulate. To minimize the arbitrariness in determining $k_f^D$, it is assumed that if the $n_B$ is $10^{3}$ times greater than the DM density, then the DM mass fraction (the ratio of DM mass to NS mass) is $\sim 1/6$~\cite{Panotopoulos:2017idn} given rise to $k_f^D\approx 33$MeV. This estimate is based on DM accumulation in the NS, ensuring that DM does not completely dominate the star's gravitational structure while still having a significant impact on the EoS. Thus, the range chosen falls within this value. The field equations at zero temperature can be found in~\cite{Lenzi:2022ypb, Das:2021hnk, Lenzi:2022ypb, Sen:2021wev, Guha:2021njn, Lourenco:2021dvh} and references therein.

Due to the relatively heavy masses of the neutralinos and their nonrelativistic nature, they are usually treated as cold DM. However, when the mass of neutralinos is reduced to the MeV scale, they may exhibit different behavior from traditional WIMPs~\cite{Ghosh:2008yh} and could be modeled as asymmetric DM, similar to the hadronic sector~\cite{Goldman:2013qla, Dutra:2018gmv, Kumar:2024zzl}. Also, light asymmetric DM can be treated as warm DM, impacting microscopic structures. Therefore, in the fixed-entropy regime, we integrated over the entire DM momentum space and varied its mass from 0.4 GeV to 0.62 GeV to study how the DM content influences the temperature distribution within the star, as well as other microscopic and macroscopic properties of PNSs. To determine the cold star configurations, we retain our approach of integrating over the full momentum space while reducing the $s_B$ to the lowest value that ensures a sufficiently low core temperature, yet preserves numerical stability. At this stage, the DM mass is varied between 1 and 2 GeV {\color{blue}as well}, as lower values in the range of 0.4–0.62 GeV do not induce significant structural changes in the star. The results of the particle distributions obtained here are then compared with those from cold and catalyzed DANSs constructed using a 200\,GeV DM particle mass, as discussed earlier, to draw relevant physical conclusions. This approach is efficient in determining the DM density and how it varies along the star's radius.  

The equation of motion 
in the mean-field approximation 
$h\rightarrow\langle h \rangle=h_0$ is given by: 
\begin{equation}
     h_0 m_h^2= {g_h}n_s^{D} + \sum_{N}{f}\frac{m_N}{v}n_b^s,
\end{equation}
where 
\begin{align}
    n_{b}^s &=\gamma_b \int \frac{d^3 k}{(2\pi)^3} \frac{m^\ast_b}{E_b} \left[f_{b\,+} + f_{b\,-}  \right],\\
     n_{D}^s &=\gamma_D \int \frac{d^3 k_D}{(2\pi)^3} \frac{m^\ast_\chi}{E_D} \left[f_{D\,+} + f_{D\,-}  \right],
\end{align}
and 
\begin{align}
    f_{b \pm}(k) &= \frac{1}{1+\exp[(E_b \mp \mu^\ast_b)/T]}, \nonumber\\
    f_{D \pm}(k_D) &= \frac{1}{1+\exp[(E_D \mp \mu_D)/T]}, \nonumber
\end{align}
$f_{b \pm}(k)$ and $f_{D \pm}(k_D)$ are the Fermi distribution functions of the nuclear and DM, respectively. While $\gamma_b =2$ is the baryon degeneracy factor, $\gamma_D=2$ is the DM degeneracy factor, and $n_b^s$ and $n_D^s$ are the scalar densities of the baryons and the DM, respectively. {$\mu_b^*$, $\mu_D$, $E_b = \sqrt{m_N^{*2}+\mu_b^{*2}}$, $E_D =\sqrt{m_\chi^{*2} + \mu_D^{2}}$ are the effective baryon chemical potential, DM chemical potential ($\mu_D$ is defined in Eq.~(\ref{mu}) below), single particle energy for baryons, and single particle energy for DM, respectively.} While $m_N^{*}$ is effective nucleon mass and $m_\chi^*$ is the effective DM mass given by: 
\begin{eqnarray}
m^*_N &=& m_N - g_{N\sigma}\sigma_0 -\sum_N f\frac{m_N}{v}h_0 ,  \nonumber \\
m_\chi^* &=& m_\chi - g_h h_0.
\end{eqnarray}
The EoS of the DANSs becomes:
\begin{align}
     \varepsilon_t={}& \varepsilon_H + \gamma_D \int \frac{d^3 k_D}{( 2\pi)^3} E_D \left [f_{D+} +f_{D-} \right]+ m_h^2h_0^2 \nonumber\\
     =& \varepsilon_H+\varepsilon_D,\\
    P_t =&P_H +  \gamma_D \int \frac{d^3 k_D}{( 2\pi)^3} \frac{k^4_D}{E_D} \left [ f_{D+} +f_{D-} \right] - m_h^2h^2_0\nonumber\\
    =& P_H +P_D,
\end{align}
where $\varepsilon_H$ and $P_H$ are the total energy density and pressure of the $\beta$-equilibrated baryonic matter and $\varepsilon_D$ and $P_D$ are the energy density and pressure of the neutral DM particles and $\varepsilon_t$ and $P_t$ are the total energy density and total pressure of the stellar system respectively. The $\mu_D$ is determined from,
\begin{equation}\label{mu}
    \mu_D = \dfrac{\partial \varepsilon_D}{\partial n_D}\Big\vert_{s_B},
\end{equation}
where 
\begin{equation}
n_D = \gamma_D \int \frac{d^3 k_D}{(2\pi)^3}  \left[f_{D\,+} - f_{D\,-}  \right],
\end{equation}
density of the DM particles. We then numerically compute Eq.~(\ref{mu}) for different evolutionary stages. The isentropic EoS of the DANS is subsequently determined using the free energy relation, $\mathcal{F} = \varepsilon_t - Ts$, where $s$ represents the total entropy density of the system. The relationship connecting $s$, $T$, $\varepsilon_t$, and $P_t$ is given by:
\begin{equation}\label{q}
sT =  P_t+\varepsilon_t  - 
\sum_b \mu_b n_b - \sum_L \mu_L n_L - \sum_D \mu_D n_D.
\end{equation}
Applying $\beta$-equilibrium and charge neutrality conditions to simplify the above expression, we obtain
\begin{equation}\label{s1}
     sT=P_t+\varepsilon_t -n_B\mu_B -n_D\mu_D,
 \end{equation}
for the neutrino-transparent regime of the star's evolution and 
\begin{equation}\label{s2}
    sT=P_t+\varepsilon_t -n_B\mu_B-\mu_{\nu_e}(n_{\nu_e} +n_e) -n_D\mu_D,
 \end{equation}
for the neutrino-trapped regime. We fixed the $s_B$ to determine the stages of the star's evolution in the form $s_B =\frac{s}{n_B}$. {In our notation, the subscript $b$ refers to the individual baryon species, specifically protons $(p)$ and neutrons $(n)$, while $B$ represents the total baryonic content of the matter. The baryon chemical potential is determined by the $\beta$-equilibrium condition, $\mu_n = \mu_p + \mu_e$, with $\mu_e = \mu_\mu$ and the overall baryon chemical potential given by $\mu_B = \mu_n$. The baryon number density is then expressed as $n_B = n_n + n_p$ (see Ref.~\cite{Menezes:2021jmw} for a detailed review on the $\beta$-equilibrium, charge neutrality, and baryon number conservation of stellar matter). 
}


\begin{table}
\caption {DDME2 parameters.}
\begin{center}
\begin{tabular}{ |c| c| c| c| c| c| c| }
\hline
 meson($i$) & $m_i(\text{MeV})$ & $a_i$ & $b_i$ & $c_i$ & $d_i$ & $g_{i N} (n_0)$\\
 \hline
 $\sigma$ & 550.1238 & 1.3881 & 1.0943 & 1.7057 & 0.4421 & 10.5396 \\  
 $\omega$ & 783 & 1.3892 & 0.9240 & 1.4062 & 0.4775 & 13.0189  \\
 $\rho$ & 763 & 0.5647 & $\cdots$ & $\cdots$ & $\cdots$ & 7.3672 \\
 \hline
\end{tabular}
\label{T}
\end{center}
\end{table}

\section{Results and Analysis}\label{r}

\begin{table*}
\centering
\scriptsize
\setlength{\tabcolsep}{5pt}
\renewcommand{\arraystretch}{1.15}
\caption{The properties of the star determined at the maximum mass ($M_{\text{max}}$), radii ($R$), baryon mass ($M_B$), central baryon density ($n_c$), central neutrino fraction ($x_{\nu_c}$) and the core temperature ($T_c$). {The core temperature at a fixed baryon mass of $1.49\,M_\odot$ is denoted by $T_{c_{1.49}}$ and its corresponding central baryon density is denoted by $n_{c_{1.49}}$.} In this table, we chose different values for $m_\chi$ in the case of the fixed entropy and cold stellar configurations and tabulate their corresponding properties. The cold star configuration corresponds to $s_B=0.2,\; Y_{\nu_e}=0$ for the DANSs and $T=0$ for matter with no DM content.}
\vspace{0.2cm}
\begin{tabular}{|l|l|l|l|l|l|l|l|l|l|}
\hline
$s_B; Y_{L,e}$ & $m_\chi$ [GeV] & $M_{\text{max}}[M_{\odot}]$ & $R$ [km] & $M_B [M_{\odot}]$ & $n_c$ [fm$^{-3}$] & $T_c$ [MeV] & $x_{\nu_c}$ & {$T_{c_{1.49}}$ [MeV]} & {$n_{c_{1.49}}$ [fm$^{-3}$]} \\ \hline\hline

$1;\; 0.4$ & 
\begin{tabular}[c]{@{}l@{}}0 \\ 0.4 \\ 0.62\end{tabular} & 
\begin{tabular}[c]{@{}l@{}}2.44 \\ 2.36 \\ 2.27\end{tabular} & 
\begin{tabular}[c]{@{}l@{}}12.22 \\ 11.86 \\ 11.42\end{tabular} & 
\begin{tabular}[c]{@{}l@{}}2.81 \\ 2.61 \\ 2.37\end{tabular} & 
\begin{tabular}[c]{@{}l@{}}0.80 \\ 0.81 \\ 0.83\end{tabular} & 
\begin{tabular}[c]{@{}l@{}}30.12 \\ 7.11 \\ 6.69\end{tabular} & 
\begin{tabular}[c]{@{}l@{}}0.0697 \\ 0.0962 \\ 0.1583\end{tabular} & 
\begin{tabular}[c]{@{}l@{}}18.31 \\ 16.15 \\ 15.23\end{tabular} &
\begin{tabular}[c]{@{}l@{}}0.32 \\ 0.35 \\ 0.38\end{tabular} \\ \hline

$2;\; 0.2$ & 
\begin{tabular}[c]{@{}l@{}}0 \\ 0.4 \\ 0.62\end{tabular} & 
\begin{tabular}[c]{@{}l@{}}2.49 \\ 2.23 \\ 2.01\end{tabular} & 
\begin{tabular}[c]{@{}l@{}}12.73 \\ 11.51 \\ 10.47\end{tabular} & 
\begin{tabular}[c]{@{}l@{}}2.89 \\ 2.23 \\ 1.77\end{tabular} & 
\begin{tabular}[c]{@{}l@{}}0.76 \\ 0.81 \\ 0.87\end{tabular} & 
\begin{tabular}[c]{@{}l@{}}71.94 \\ 15.58 \\ 14.27\end{tabular} & 
\begin{tabular}[c]{@{}l@{}}0.0099 \\ 0.3639 \\ 0.4589\end{tabular} & 
\begin{tabular}[c]{@{}l@{}}42.52 \\ 33.87 \\ 35.97\end{tabular} &
\begin{tabular}[c]{@{}l@{}}0.28 \\ 0.37 \\ 0.51\end{tabular} \\ \hline

$2;\; Y_{\nu e}=0$ & 
\begin{tabular}[c]{@{}l@{}}0 \\ 0.4 \\ 0.62\end{tabular} & 
\begin{tabular}[c]{@{}l@{}}2.49 \\ 2.23 \\ 2.01\end{tabular} & 
\begin{tabular}[c]{@{}l@{}}12.72 \\ 11.41 \\ 10.36\end{tabular} & 
\begin{tabular}[c]{@{}l@{}}2.91 \\ 2.22 \\ 1.73\end{tabular} & 
\begin{tabular}[c]{@{}l@{}}0.76 \\ 0.81 \\ 0.88\end{tabular} & 
\begin{tabular}[c]{@{}l@{}}70.00 \\ 15.59 \\ 14.27\end{tabular} & 
\begin{tabular}[c]{@{}l@{}}$\cdots$ \\ $\cdots$ \\ $\cdots$\end{tabular} &
\begin{tabular}[c]{@{}l@{}}41.95 \\ 34.90 \\ 37.54\end{tabular} &
\begin{tabular}[c]{@{}l@{}}0.29 \\ 0.37 \\ 0.54\end{tabular} \\ \hline

Cold Stars & 
\begin{tabular}[c]{@{}l@{}}0 \\ 1 \\ 2\end{tabular} & 
\begin{tabular}[c]{@{}l@{}}2.49 \\ 2.42 \\ 2.27\end{tabular} & 
\begin{tabular}[c]{@{}l@{}}12.05 \\ 11.83 \\ 11.13\end{tabular} & 
\begin{tabular}[c]{@{}l@{}}3.01 \\ 2.85 \\ 2.44\end{tabular} & 
\begin{tabular}[c]{@{}l@{}}0.81 \\ 0.83 \\ 0.87\end{tabular} & 
\begin{tabular}[c]{@{}l@{}}0.00 \\ 5.67 \\ 4.40\end{tabular} & 
\begin{tabular}[c]{@{}l@{}}$\cdots$ \\ $\cdots$ \\ $\cdots$\end{tabular} &
\begin{tabular}[c]{@{}l@{}}$\cdots$ \\ $\cdots$ \\ $\cdots$\end{tabular} &
\begin{tabular}[c]{@{}l@{}}$\cdots$ \\ $\cdots$ \\ $\cdots$\end{tabular} \\ \hline

\end{tabular}
\label{ma}
\end{table*}

\begin{figure*}
  \includegraphics[scale=0.6]{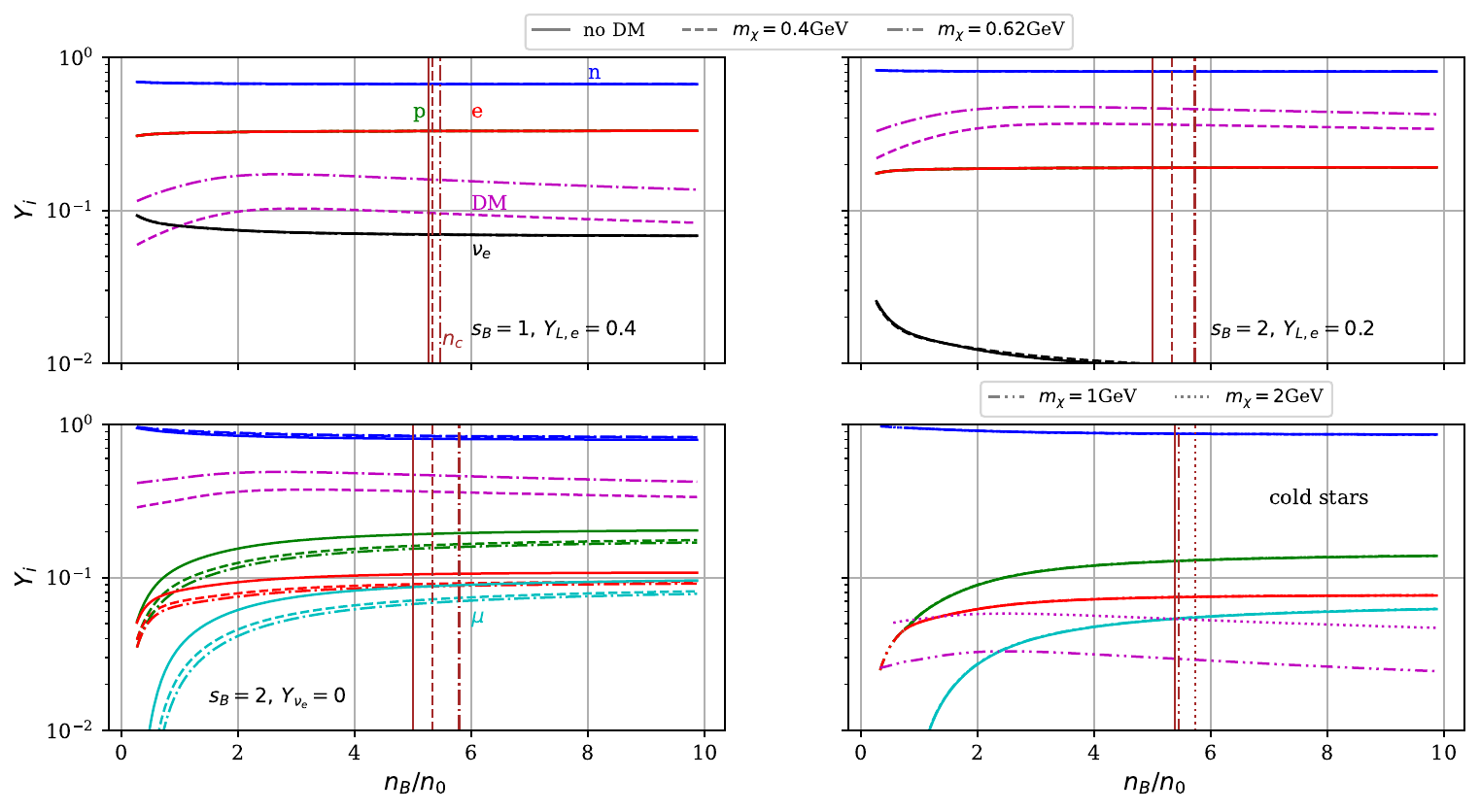}
\caption{The figure shows the particle fraction as a function of $n_B/n_0$. We show four snapshots comprising a neutrino-trapped regime (upper panels) and a neutrino-transparent regime (bottom panels). Stellar matter without a DM component is shown with solid lines across all four evolutionary stages.  For fixed-entropy configurations, dashed lines correspond to stars admixed with DM of mass $m_\chi = 0.4\,\text{GeV}$, while dash-dot-dash lines indicate $m_\chi = 0.62\,\text{GeV}$. In the cold stars case, dash-double-dot lines represent stars admixed with 1 GeV DM, and dotted lines correspond to those mixed with 2 GeV DM mass. The brown vertical lines indicated in the figures represent the position of the central baryon density, $n_c$, of the corresponding stellar configuration, following the same line style.
}
    \label{pf}
\end{figure*}

\begin{figure}
    \centering
    \includegraphics[scale=1]{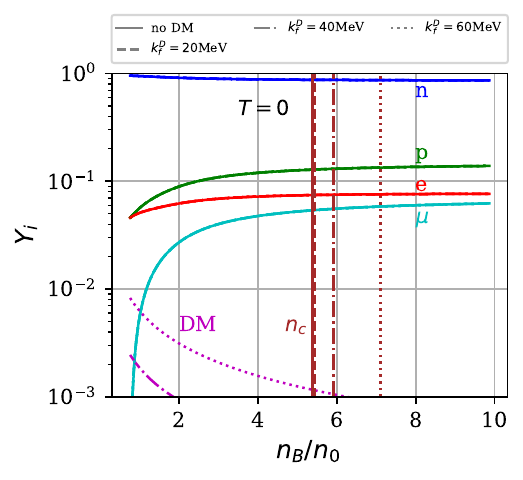}
    \caption{The particle fraction for a neutralino DM particle with mass 200\,GeV admixed with NSs is shown. 
    In this analysis, the DM mass is fixed, and the DM content is controlled by varying the DM Fermi momentum ($k_F^D$). Different line styles represent the values of $k_F^D$ considered: the solid line corresponds to $k_F^D = 0$, the dashed line to $k_F^D = 20\,\text{MeV}$, the dash-dot line to $k_F^D = 40\,\text{MeV}$, and the dotted line to $k_F^D = 60\,\text{MeV}$. For readers interested in the corresponding stellar structure associated with these DM configurations, further details can be found in Ref.~\cite{Das:2021hnk, Lopes:2023uxi, Das:2018frc, Lopes:2024ixl, Lourenco:2021dvh}.}
    \label{pf1}
\end{figure}

We explore the impact of the neutralino DM candidate on the evolution of PNSs through a series of snapshots. The DM is considered a potential candidate for direct detection and collider experiments. {It is the lightest supersymmetric DM with a mass in the GeV–TeV range; we specifically consider a mass of 200\,GeV while varying the DM Fermi momentum to compare with results for the final stage of the star's evolution in our approach, when the star is in a cold configuration.} This choice is reasonable as it satisfies direct detection and collider experimental constraints. However, assuming such a high mass in a supernova remnant implies that neutralinos would decouple from the thermal plasma early, reducing their influence on the temperature evolution. In this case, their weak interactions may prevent them from efficiently contributing to heating mechanisms, such as neutrino diffusion and shock wave propagation. This effect contrasts with the conventional expectation that these processes heat supernova remnants after the explosion. 

Therefore, to study PNSs, we reduce the mass of neutralinos to about the GeV scale, integrating over the entire momentum space and varying their mass between 0.4\,GeV and 0.62\,GeV for PNS and 1\,GeV and 2\,GeV for cold stars. This choice is constrained by the $2\rm M_\odot$ NS mass limit.  For $ m_\chi> 0.62$ GeV, we obtained lower-mass PNSs less than the required 2$\rm M_\odot$ limit for the NSs, which we discarded from our PNS analysis. {The DM mass limit $m_\chi \leq 0.62\,\mathrm{GeV}$ is specific to the DDME2 EoS satisfying the 2$\rm M_\odot$ constraint and may vary with EoS stiffness and DM interaction strength; exploring this is left for future work.} Although low-mass neutralinos are unconventional in minimal supersymmetric standard models, they have been explored in extended supersymmetric frameworks and other beyond-the-standard-model scenarios. For instance, in~\cite{Dreiner:2003wh}, the authors investigated the parameter space of supersymmetric models that accommodate light neutralinos ($m_\chi \lesssim 1$ GeV), focusing on constraints from the supernova event SN1987A~\cite{Kamiokande-II:1987idp}. Light neutralino DM in the next-to-minimal supersymmetric standard model is discussed in~\cite{PhysRevD.73.015011}, while cosmological constraints on light DM candidates can be found in~\cite{Deng:2023twb}. A review of the impact of varying DM mass on the macroscopic properties of an NS can be found in~\cite{Hajkarim:2024ecp} and the references therein. 

\subsection{Constraints from PNS}

 To investigate the behavior of stars at the final stage of their evolution, when they are considered cold and catalyzed, we retain our approach of fixing only the DM mass and integrating over the full momentum space, as adopted for the evolution of PNSs. This requires keeping $s_B$ as low as possible, since our numerical codes do not converge for $s_B = 0$.
Therefore, to approximate cold conditions while preserving both our methodology and numerical stability, we set $s_B = 0.2$. Although this implies that the core temperature is not strictly zero or less than 1 MeV, as expected for cold-catalyzed stars, it remains relatively low, in the range of 4.40 to 5.67 MeV, as shown in Tab.~\ref{ma}. Indeed, in Ref.~\cite{Largani:2023kjx}, the authors compared the microscopic and macroscopic properties of neutrino-transparent, $\beta$-equilibrated NSs at $T = 0$ with those at $s_B = 0.1$ and $s_B = 0.5$, and found that the deviation in stellar properties is on the order of 1\%. 

In our case, we also calculated the mass sequence of the star with no DM content for $s_B =0.2,\; Y_{\nu_e}=0$ and determined a maximum mass of 2.49 $M_\odot$ with radius 12.05 km, which is identical to the $T=0$ data on Tab.~\ref{ma} without a difference. Thus, fixing a low $s_B$ allows for a balance between the physical realism of $s_B = 0$ and computational feasibility. This approximation is advantageous, as it avoids the need to fix the DM number density $n_D$, which would necessitate treating the DM Fermi momentum as a constant--a common but simplifying assumption in the literature (See Ref.~\cite{Das:2021hnk, Lopes:2024ixl} and references therein). Moreover, as we shall see below, it facilitates a more natural comparison with previous studies, particularly those employing a 200\,GeV neutralino mass to investigate DM-admixed neutron stars. It is important to note that the cold star configuration corresponds to $T = 0$ for stars without a DM component, and $s_B = 0.2,\; Y_{\nu_e} =0$ for DANSs. These respective conditions will be collectively referred to as ``cold stars" here and throughout the remainder of this work.

In Tab.~\ref{ma}, we observe that an increase in the DM content, through the variation of $m_\chi$, leads to a decrease in $M_{\rm max}$ and $M_B$ along with a significant reduction in radius $R$, resulting in more compact DANSs. This increased compactness is due to the enhanced gravitational attraction from DM, which compresses baryonic matter and reduces the baryonic mass required for stable stellar configurations. The $n_c$ increases when the DM content increases due to the enhanced compactness of the star. In contrast, $T_c$ decreases significantly as the DM content increases, indicating that DM absorbs thermal energy from the stellar matter without efficiently re-emitting it. This behavior is consistent with the discussion in Fig.~\ref{tp}, where we examined how the temperature varies with $n_B/n_0$  at different DM contents. However, a slightly different trend emerges when the baryon mass  $M_B$ is held constant. In the first stage, we observe a systematic decrease in $T_{c_{1.49}}$. In the second and third stages, $T_{c_{1.49}}$ remains lower than in the no-DM case, supporting the interpretation of thermal energy absorption by DM. Interestingly, for $m_{\chi} = 0.62~\mathrm{GeV}$, a slight reheating effect is observed at fixed $M_B$. This arises from adiabatic compression: the heavier DM increases the $n_{c_{1.49}}$, which locally elevates $T_{c_{1.49}}$ through the relation $T \propto n_B^{2/3}$ \cite{Shapiro1983, Glendenning2000}. This localized reheating does not contradict the overall net cooling role of DM, as clearly shown in Fig.~\ref{tp}, but rather highlights the impact of gravitational restructuring of the temperature profile. 

The increase in the \(x_{\nu_c}\) with increasing \(m_\chi\) suggests that DM influences weak interactions, thermal evolution, and consequently stellar compactness. As DM absorbs thermal energy (observed from decreasing $T_c$ with increasing $m_\chi$) from the stellar matter, it can alter weak interaction rates, enhancing neutrino production while reducing their escape, leading to a higher \(x_{\nu_c}\). Additionally, the increase in \(n_c\) due to the presence of DM further enhances neutrino trapping, as the neutrino mean free path decreases in denser regions.

This work assumes a non-annihilating fermionic DM candidate \cite{Petraki:2013wwa} that is already admixed at the onset of PNS evolution. We do not model its dynamical accumulation through capture, fallback accretion, or annihilation. The heat reservoir role attributed to DM in our study arises purely from gravitational restructuring, such as increased central densities and deeper gravitational potentials, and not from annihilation or kinetic heating. This distinction is especially relevant for asymmetric DM scenarios, where the DM number is conserved, as considered here. For sub-GeV DM particles, typical accretion timescales exceed the $\sim$10 s thermal evolution timescale of the PNS \cite{Pons:1998mm}, supporting our assumption of early-time admixture. Additionally, stars forming in dense environments, such as galactic centers, may encounter ambient DM densities sufficient to enable significant pre-collapse accumulation \cite{Gnedin:2003rj}. While our simplified setup isolates the gravitational and thermodynamic influence of an admixed DM component, a more complete treatment, especially for light, self-annihilating, or self-interacting DM, would require time-dependent modeling that accounts for capture dynamics, annihilation, progenitor history, and post-bounce evolution.

Additionally, the table shows that increasing the DM particle mass $m_\chi$ does not lead to drastic changes in the maximum mass as the star cools toward a cold, catalyzed configuration. 
The choice of the higher DM masses at this stage is motivated by our observation that the stellar structure is not significantly affected when varying $m_\chi$ between 0.4 and 0.62 GeV during the cooling process. For instance, we obtained maximum masses of 2.48 and 2.47 $M_\odot$ for $m_\chi = 0.4$ and 0.62 GeV, respectively, values that are nearly identical to the maximum mass calculated at $T = 0$ for a stellar configuration with no DM content. This indicates that within this DM mass range, the impact on the stellar structure during the transition to the cold-catalyzed stage is minimal. Consequently, we chose $m_\chi = 1$ and 2 GeV at this stage, for which the maximum mass remains above the 2 $M_\odot$ threshold for NSs but shows an identifiable change in the stellar structure.  In contrast, in the case of PNSs, increasing the DM mass to 0.64 GeV causes the maximum mass at the third stage to drop to 1.99 $M_\odot$, falling just below the observationally supported threshold used as a baseline in this study. %

\subsection{EoS properties of OM--DM  model}

In Fig.~\ref{pf}, we present the particle fractions ($Y_i$) distributed in the stellar matter as a function of $n_B/n_0$. $Y_i$ is governed by the expression;
\begin{equation}
    Y_i = \dfrac{n_i}{n_B},
\end{equation}
$n_i$ represents the density of the individual particle number, including the DM particles. Given that the fraction of DM particles relative to nucleons is defined as, $F_{DM}=N_{DM}/N_N$, where $N_{DM}$ is the number of DM particles and $N_N$ is the number of nucleons, we can obtain $F_{DM}$ from $Y_i$ since DM and OM particles occupy the same volume under the single fluid framework~\cite{Hajkarim:2024ecp}.  Along the figure, from the first stage (top left) to the final stage (bottom right), we observe that the isospin asymmetry $\delta = (n_n-n_p)/(n_n+n_p)$ -- where $n_n$ is the neutron number density and $n_p$ is the proton number density, respectively -- increases across the panels. The $\delta$ is directly related to the nuclear symmetry energy, which quantifies the energy cost associated with an imbalance between neutron and proton numbers in the EoS. 

We start by pointing out some general features seen in Fig.~\ref{pf} relevant for the EoS of our model of OM and DM particles. In all panels, we observe that increasing the DM particle mass raises the ratio \( Y_{DM} = n_{DM} / n_B \). This behavior results from the decreasing contribution of the DM kinetic energy, which becomes dominated by the attractive self-interaction mediated by the Higgs boson. This trend arises from the minimization of the mean-field grand-canonical potential, which determines the equations of motion and the mean-field EoS. 
Furthermore, at low baryonic densities, the system tends to minimize the presence of DM particles, as it is energetically more favorable to reduce their number. This is because the attractive interaction among visible baryons dominates the grand-canonical potential in this regime. However, as \( n_B \) increases, the repulsive core of the nuclear interaction becomes significant, reducing the contribution from baryon-baryon attraction. As a result, the attractive self-interaction among DM particles becomes increasingly relevant, allowing \( Y_{DM} \) to grow.

In the top panels (neutrino-trapped regime), we observe that in the first stage, the maximum amount of DM accumulation is reached when $m_\chi=0.62$ GeV, representing approximately $14\%$ of the matter content. In the second stage (top right), during neutrino diffusion, the DM content increases to about $32\%$ and shows noticeable changes in the neutrino emission curves. The lower DM accumulation in the first stage compared to the second one can be attributed to the higher $Y_{L,e}$ which leads to an increase in lepton interaction, on the other hand, a lower $Y_{L,e}$ enhances DM accumulation due to an increase in baryonic interactions~\cite{Busoni:2021zoe}. The delay in the neutrino emission (more obvious in the second stage than the first stage) due to the presence of DM 
alters the thermal evolution of the star because the star cools down mainly through neutrino diffusion. This could affect the conventional way of determining the star's age and serve as a potential observational signature for DM in NS.

In the bottom panels (neutrino-transparent regime), we observe a higher DM content of about 40\% in the third stage (bottom left), while in the last stage (bottom right), the DM composition is about 5\%. In the third stage, all neutrinos have escaped from the stellar core, and the matter has reached its peak temperature before beginning to cool. DM influences the star's composition at this stage more significantly by enhancing neutron production while suppressing proton and lepton production. This effect may result from modifications to beta equilibrium induced by DM due to its 
 interactions with baryons (see the difference between Eqs.~(\ref{s1}) and (\ref{s2})). The changes in particle distributions caused by the presence of DM can impact radiative properties, such as X-ray emissions and thermal spectra, as well as pulsar timing. These effects may result in slow or rapid rotational changes induced by alterations in the star's internal dynamics, which pulsar timing arrays could detect. The final stage corresponds to the cold configuration of the star. At this point, the DM content is significantly reduced compared to earlier stages, despite the higher DM particle mass. As a result, the impact on the overall particle distribution is minimal. This diminished influence may be attributed to the weaker interactions between DM and OM as the stellar matter cools (this is clearer in Fig.~\ref{cpl1} when the interaction strength between the DM and the OM was investigated).

In Fig.~\ref{pf1}, we present the particle distribution in a cold and catalyzed star admixed with a 200\,GeV neutralino DM particle, for comparison with our results at the final stage of stellar evolution. The figure illustrates that as the star cools, it can accommodate a higher DM mass without significant alterations in the particle composition, despite visible changes in the stellar structure for different DM contents set via the Fermi momentum $k_f^D$. Since several authors have already examined the structure of DANSs using similar models with the same DM mass under comparable conditions, we do not repeat those calculations here (see Refs.~\cite{Lopes:2024ixl, Lourenco:2022fmf, Das:2021hnk} for detailed discussions). These results support the view that DM accreted over time in an NS can remain gravitationally bound without disrupting the star’s equilibrium configuration, although it may reduce the mass-radius ratio and increase compactness, as discussed in~\cite{Kumar:2025ytm}. A similar conclusion was reached in~\cite{DelPopolo:2020hel}, where it was shown that equilibrium is preserved during DM accretion unless a critical DM mass threshold is exceeded. 

Even though such a heavy DM candidate is not excluded by current NS observations, the similarity in particle fraction profiles across different DM contents indicates that astrophysical data alone may be insufficient to place stringent constraints on DM properties. This highlights the need to investigate more sensitive observables, such as temperature distributions, neutrino emission profiles, or supernova remnant signatures, to gain deeper insights into the role and behavior of DM in compact stars. This result is comparable with the one obtained for the last stage of the stellar evolution presented in Fig.~\ref{pf} representing the cold stellar configuration. Here, the different DM masses do not show any difference in the particle distribution plot, even though Tab.~\ref{ma} shows their structural differences. It is important to note that for a fixed DM Fermi momentum, the DM fraction $Y_{\text{DM}}$ shows a decreasing trend with increasing $n_B/n_0$, in contrast to the behavior observed in Fig.~\ref{pf}. This marked difference arises because fixing the Fermi momentum freezes both the DM repulsive pressure and its number density. As the baryon density grows, the DM fraction $Y_{\text{DM}}$ naturally decreases.

\begin{figure}[t!]
  \includegraphics[scale=1]{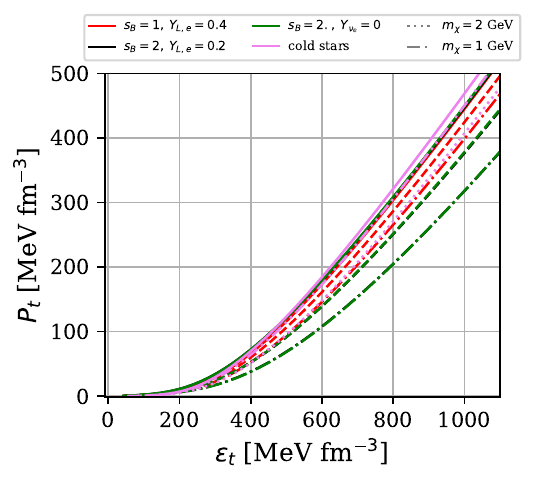}
 \caption{This figure displays pressure as a function of energy density. Stellar matter without a DM component is shown with solid lines across all four evolutionary stages.  For fixed-entropy configurations, dashed lines correspond to stars admixed with DM of mass $m_\chi = 0.4\,\text{GeV}$, while dash-dot-dash lines indicate $m_\chi = 0.62\,\text{GeV}$. In the cold stars case, dash-double-dot lines represent stars admixed with 1 GeV DM, and dotted lines correspond to those mixed with 2 GeV DM mass. 
}
    \label{eos}
\end{figure}

In Fig.~\ref{eos}, we present the total pressure as a function of total energy density, showing that the presence of DM leads to a softening of the EoS. As the DM content increases through increasing DM mass, the EoS becomes progressively softer. In the model framework, where DM and OM mix, the total pressure $P_t$ and the total energy density $\varepsilon_t$ account for both components. DM primarily enhances gravitational attraction by increasing the energy density while contributing minimally to pressure. Due to its weak interaction with OM, DM effectively increases the total mass without providing significant structural support. This results in a more compact star and a lower maximum mass (as we shall see later in Fig.~\ref{m-r}), a characteristic feature of softer EoSs. Further discussions on the effect of DM on NS EoS can be found in~\cite{Lenzi:2022ypb, Das:2018frc, Das:2020vng} and references therein. The EoS of the second and the third stages show an overlap in the above figures, but the differences between them become clearer in Fig.~\ref{pf} and Figs.~\ref{tp}, \ref{m-r}, and \ref{cpl} below, and the other plots related to the DM properties.

\begin{figure}
  \centering
  \includegraphics[scale=1]{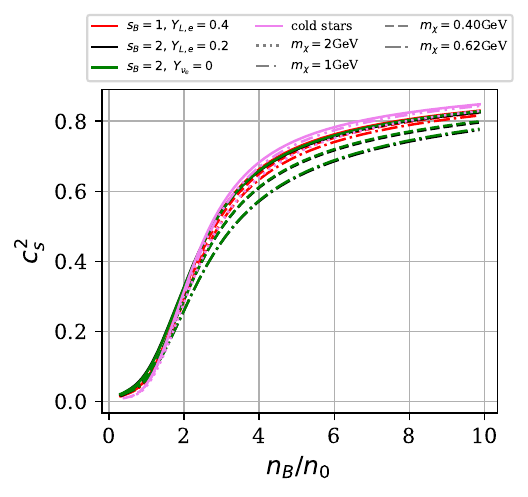}
 \caption{Variation of the squared speed of sound, $c_s^2$, with $n_B/n_0$ for different DM mass parameterizations.}
    \label{ss}
\end{figure}

    The squared speed of sound ($c_s^2$) in Fig.~\ref{ss} is obtained from the EoS via
\[
c_s^2 = \frac{dP}{d\varepsilon},
\]
serving as a macroscopic thermodynamic measure of the EoS stiffness. In exactly conformal matter, $c_s^2 = {1}/{3}$  (with $c=1$) and approaches this limit from below in high-density quark matter \cite{Kurkela:2009gj}, while at densities below nuclear saturation, $c_s^2 \ll 1$. Chiral EFT calculations indicate that in hadronic matter, $c_s^2$ can exceed the conformal value, reaching $c_s^2 \gtrsim 0.5 $ \cite{Bedaque:2014sqa, Leonhardt:2019fua}. The speed of sound is further bounded by the causality constraint, $c_s^2 \leq 1$, and by the thermodynamic stability condition, $c_s^2 > 0$. Our results satisfy the causality constraint, with all EoS parameterizations remaining below $c_s^2 < 1 $. We observe that $c_s^2$ decreases with increasing DM mass in all PNS evolutionary stages analyzed, mainly because the presence of DM softens the EoS. The highest c$_s^2$ is achieved when the star is cold and more compact, with higher $n_c$, followed by the first stage of PNS evolution when neutrinos are trapped. The intermediate (second) stage, when the star is deleptonizing, and the third (neutrino-transparent) stage, when the star is hot and expanded, show lower values of  $c_s^2$ due to the thermal softening of the EoS.

\begin{figure}
  \includegraphics[scale=1]{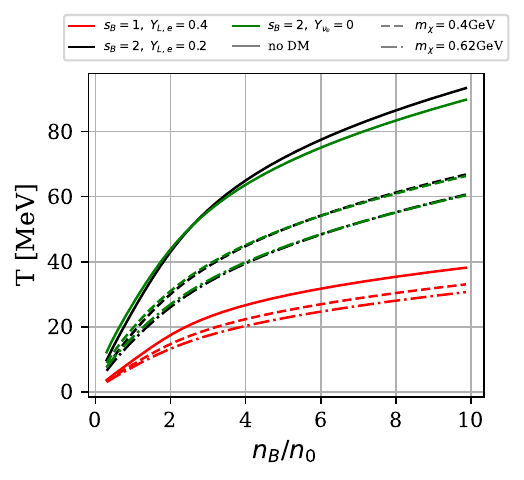}
\caption{The figure shows the temperature profiles within the stellar matter from a newly born star as it evolves through neutrino diffusion to the neutrino-transparent stage.  
}
    \label{tp}
\end{figure}
Fig.~\ref{tp} represents the temperature distributions in the stellar matter as a function of $n_B/n_0$. In general, the temperature increases with increasing $s_B$ and decreasing $Y_{L,e}$ following the expected behavior of NS evolution~\cite{Prakash:1996xs}. In the first stage, the temperature remains relatively low when the matter has a higher $Y_{L,e}=0.4$ and lower $s_B = 1$.  In the second stage, as neutrino diffusion progresses, the lepton fraction decreases $Y_{L,e}=0.2$. At the same time, the entropy increases to $s_B=2$, leading to a rise in temperature, which reaches its maximum in the core. In the third stage, after neutrinos have fully escaped from the stellar core, the matter is heated to its peak before cooling begins. 
During this stage, the temperature remains highest at lower densities until reaching $\sim 3n_0$ after which it gradually decreases toward the core. This behavior follows the general thermal evolution pattern of PNSs (see Ref.~\cite{Issifu:2023qyi} and references therein).

\subsection{Stellar Structure}
When the DM component is introduced into the matter, it reduces the temperature distribution below the threshold observed in the OM case. The magnitude of the temperature decreases depending on the amount of DM in the star. A higher DM content leads to lower temperature profiles, whereas a lower DM content results in relatively higher temperature profiles. This suggests that DM absorbs thermal energy without efficiently re-emitting it, thereby affecting neutrino emission dynamics (as discussed in Fig.~\ref{pf}) and altering the thermal energy distribution within the stellar matter. As shown in Eqs.~(\ref{q}) to (\ref{s2}), DM modifies the EoS of OM, leading to changes in the overall thermal structure of the star. These effects can influence the star’s evolutionary history and impact its observable properties, as a lower thermal photon emission rate from the star’s surface is likely a result of the reduced temperature of the stellar matter (see Ref.~\cite{Kouvaris:2007ay} on the impact of DM on NS cooling).

\begin{figure}
  \includegraphics[scale=0.9]{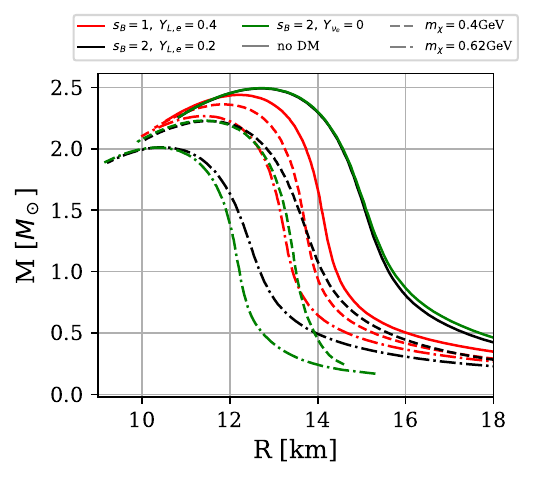}
   \includegraphics[scale=0.9]{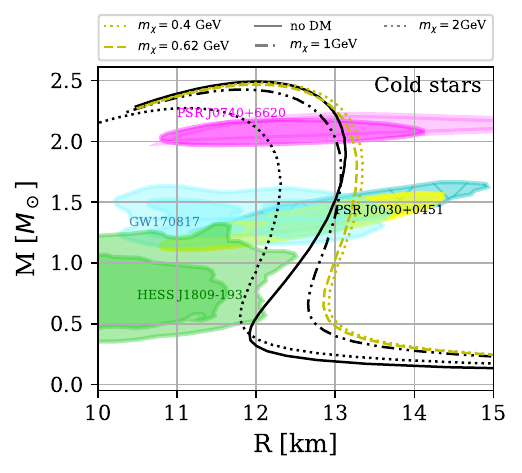}
\caption{The figure displays the gravitational mass of the sequence of stars as a function of radius. The upper panel displays the fixed-entropy stars, which comprise both neutrino-trapped and neutrino-transparent regimes. The bottom panel displays the star's structure when it is cold. Observational data shown on the plot: magenta represents PSR J0740+60.62~\cite{riley2021, Miller:2021qha}, yellow and cyan represent PSR J0030+0451~\cite{riley2019, Miller:2019cac}, light blue represents the gravitational wave event GW170817~\cite{LIGOScientific:2017vwq, LIGOScientific:2018cki}, and light green contours represent HESS J1731-347~\cite{Klochkov:2014ola}. Stars without a DM component are shown with solid lines across all four evolutionary stages. For fixed-entropy configurations, dashed lines correspond to stars admixed with DM of mass $m_\chi = 0.4\,\text{GeV}$ (doted lines) and $m_\chi = 0.62\,\text{GeV}$, while dash-dot-dash lines indicate $m_\chi = 0.62\,\text{GeV}$. In the cold star case, the dotted yellow curves represent $m_\chi = 0.4\,\text{GeV}$, while the dashed yellow lines correspond to $m_\chi = 0.62\,\text{GeV}$. The black line styles are identified as follows: the solid black line represents the case without DM, the dashed-double-dot lines indicate stars admixed with 1 GeV DM, and the dashed lines correspond to those with 2 GeV DM.
}
    \label{m-r}
\end{figure}

Fig.~\ref {m-r} shows the mass-radius relation of the stars. The upper panel represents stars with fixed entropy, while the lower panel represents the cold NS configuration. From the upper panel, we observe that the radius of the star generally increases with increasing entropy and decreasing lepton fraction~\cite{Janka:2006fh, Raduta:2020fdn}. Notably, the mass-radius diagram reveals a clearer distinction between the structures of stars in the second and third stages than what can be inferred from the EoS (Fig.~\ref{eos}), where the curves exhibit overlap. The effect of DM is observed in the relative shrinking of the curves as $m_\chi$ increases. This behavior arises because DM contributes significantly to the total energy density of the star while providing relatively little pressure to counteract gravitational collapse, {as the self-interaction is attractive and only kinetic energy accounts for that.} As a result, the increased gravitational attraction pulls the baryons inward, leading to a more compact star with a smaller mass and radius. Comparing the structures of stars in the second and third stages of evolution, we deduce that the impact of DM becomes more significant after neutrinos have fully escaped from the core. This is evident in the greater reduction of the stellar radius in the third stage compared to the second when DM is introduced. This can be attributed to the absence of degenerate pressure contributed by the neutrinos to support the star at this stage, making the DM effect more pronounced.

In the lower panel, we present the mass-radius curves for the cold NSs. Recent data from the NICER X-ray observatory, along with observations from binary NS mergers, have imposed stringent constraints on mass-radius relations, which any viable modern NS EoS must satisfy. In this panel, we overlay the confidence contours corresponding to several of these key observations. The two independent measurements of PSR J0740+6620, which report notably different radii, are shown in magenta: the inner contour corresponds to the analysis by Riley {\it et al.}~\cite{riley2021}, while the outer contour is from Miller {\it et al.}~\cite{Miller:2021qha}. Similarly, the results for PSR J0030+0451 are shown in yellow and cyan, representing the analyses by Riley et al.~\cite{riley2019} and Miller {\it et al.}~\cite{Miller:2019cac}, respectively. The GW170817 merger event is displayed with a light blue contour; the outer region reflects the 90\% CL constraint~\cite{LIGOScientific:2017vwq, LIGOScientific:2018cki}, while the inner region shows the 50\% CL~\cite{LIGOScientific:2017vwq}. The supernova remnant HESS J1731$-$347 is indicated in light green, with the inner and outer contours corresponding to the 68\% and 95\% CL measurements, respectively~\cite{Klochkov:2014ola}.

The curve corresponding to an NS with no DM content, as well as the ones admixed with 1 GeV DM, 0.4 GeV, and 0.62 GeV, satisfy all the displayed confidence contours except for that of the HESS data (right panel of Fig.~\ref{m-r}). In contrast, the configuration admixed with 2 GeV DM satisfies all the contours, including the HESS constraint at the 90\% CL. Cold NSs are more capable of supporting higher DM masses because, as PNSs cool, their gravitational binding energy increases, the core density profile steepens, and thermal pressure decreases. These changes result in a more compact and stable stellar structure. The increased compactness enhances the gravitational potential, allowing the star to bind additional DM without compromising its equilibrium, making cold configurations more favorable for hosting heavier DM components. It is worth noting that the heavier DM masses considered here (1 GeV and 2 GeV) are not meant to represent a continuous thermal evolution from the PNS phase. Instead, they serve to test the robustness of the stellar structure under more extreme DM conditions. However, the DM content in the star could increase through late-stage accumulation mechanisms, such as fallback accretion following a supernova, capture from dense DM environments like the Galactic center, or accretion in compact binary systems \cite{Bramante:2023djs, Bramante:2017xlb, Gnedin:2003rj}. Although these scenarios are not modeled in the present study, they provide a strong motivation to explore the potential structural impact of such DM admixtures.

\begin{figure}
  \includegraphics[scale=1]{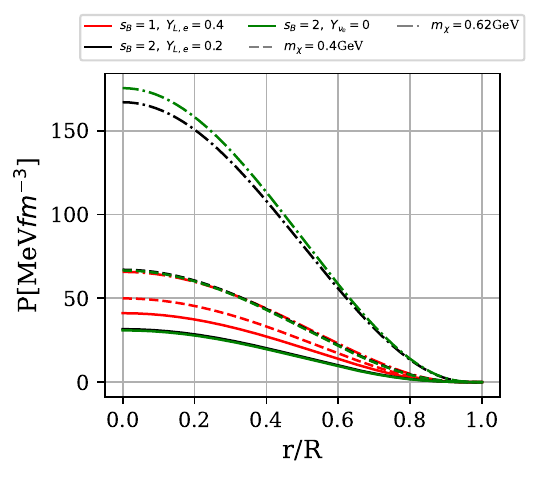}  
  \includegraphics[scale=1]{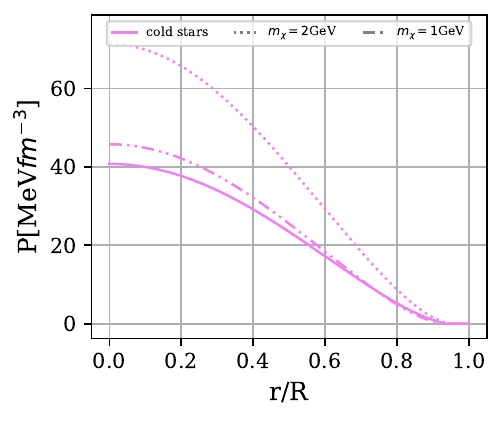}
\caption{The pressure profiles for stars with a fixed baryon mass ($M_B = 1.49\, \rm M_\odot$) corresponding to a gravitational mass of \(1.4\, \rm M_\odot \). 
Upper panel shows stellar matter with fixed-entropy configurations: OM only (solid line), OM-DM stars with $m_\chi = 0.4\,\text{GeV}$ (dashed lines) and $m_\chi = 0.62\,\text{GeV}$ (dot-dashed lines) for each case of the pair $\{s_B,Y_{L,e}\}$ equals to $\{1,0.4\}$ (red), $\{2,0.2\}$ (black) and $\{2,0\}$ (green). 
Bottom panel shows cold stellar configurations: OM only (solid line), OM-DM stars with $m_\chi = 1\,\text{GeV}$ (dash-double-dotted line) and $m_\chi = 2\,\text{GeV}$ (dotted line). }
    \label{cpl}
\end{figure}

In Fig.~\ref{cpl}, we track the evolution of a star with a fixed baryonic mass $M_B = 1.49\,\rm M_\odot$, corresponding to a gravitational mass of $1.4\, \rm M_\odot$, from birth to maturity~\cite{1986ApJ...307..178B}. Fixing $M_B$ enables us to follow the same stellar object throughout its evolution and assess the impact of DM on its core pressure. {In the upper panel, when the star is still young and lepton-rich, the presence of DM has a minimal effect on its core pressure compared to other stages. This is expected, as the high lepton and neutrino content dominates the pressure support at this stage. However, as the star enters the deleptonization phase and begins to lose neutrinos, the contribution of DM becomes increasingly significant. The impact of DM on the core pressure reaches its maximum when the star becomes neutrino-transparent.} As the star cools and contracts, the rising central density leads to an increase in core pressure during the final cold NS stage (bottom panel). It is important to note that at this point, thermal pressure no longer plays a significant role, so the absolute magnitude of the core pressure remains comparatively low. Nevertheless, DM continues to contribute to the pressure enhancement by increasing the star’s compactness, a trend that becomes more pronounced with higher DM content. Interestingly, our results show that the DM effect is most noticeable during the deleptonization phase, when the reduction in neutrino pressure allows the gravitational influence of DM to become dominant~\cite{Issifu:2023qoo, Issifu:2024fuw}.

\begin{figure}
  \includegraphics[scale=1]{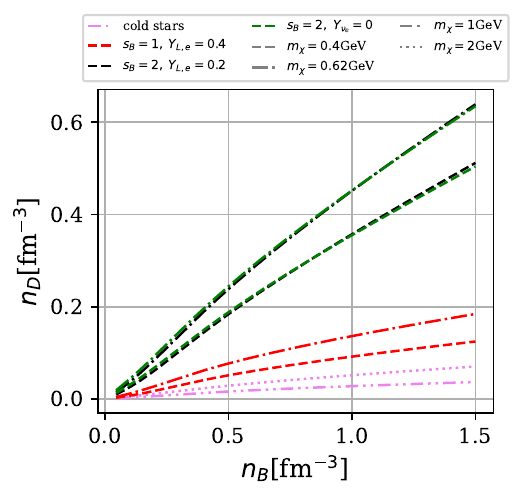}
  \includegraphics[scale=1]{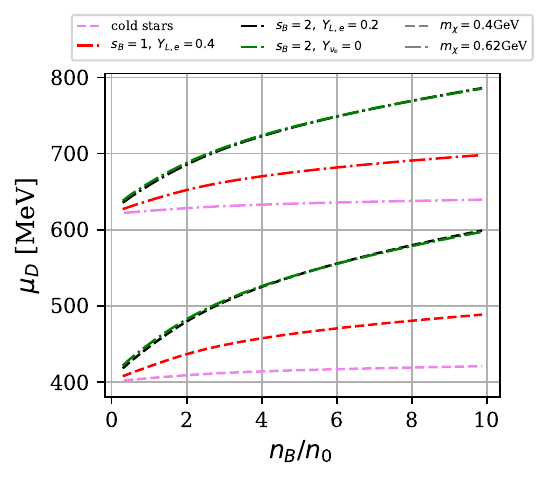}
\caption{The plots display  DM density and chemical potential against OM density. Upper panel: illustrates the relationship between OM and DM densities for cold stars (purple) with $m_\chi = 1\,\text{GeV}$ (dot-dashed line) and $m_\chi = 2\,\text{GeV}$ (dotted line), and for hot stars with $m_\chi = 0.4\,\text{GeV}$ (dashed lines) and $m_\chi =0.62\,\text{GeV}$ (dash-dotted lines) in each case of the pair $\{s_B,Y_{L,e}\}$ equals to $\{1,0.4\}$ (red), $\{2,0.2\}$ (black) and $\{2,0\}$ (green).
Lower panel shows the variation of $n_B/n_0$ with the DM chemical potential ($\mu_D$) for cold stars (purple) with $m_\chi = 1\,\text{GeV}$ (dashed line) and $m_\chi = 2\,\text{GeV}$ (dash-dotted  lines), and  for hot stars with $m_\chi = 0.4\,\text{GeV}$ (dashed lines) and $m_\chi =0.62\,\text{GeV}$ (dash-dotted  lines) in each case of  the pair $\{s_B,Y_{L,e}\}$ equals to $\{1,0.4\}$ (red), $\{2,0.2\}$ (black) and $\{2,0\}$ (green) .}
    \label{Vden}
\end{figure}

\subsection{Further EoS properties of OM-DM model }

\subsubsection{DM density and chemical potential}
In Fig.~\ref{Vden}, we show the relationship between $n_B$ and $n_D$ in the upper panel and $n_B/n_0$ with $\mu_D$ in the lower panel. The single-fluid model under consideration incorporates the Higgs portal mechanism, which allows DM to interact with OM through the Higgs boson. An increase in $n_B$ alongside $n_D$ implies a significant enhancement in the interaction between these two types of matter, which can lead to modifications in the EoS. The interplay between the two mixed fluids can significantly influence various physical properties, including particle and temperature distributions, pressure profile, and observable characteristics such as stellar structure, as illustrated in Figs. \ref{pf}, \ref{tp}, \ref{cpl}, and \ref{m-r}, respectively. The enhanced interaction between the two fluids, resulting from increased density, will facilitate rapid energy exchange between the  OM and DM, thereby promoting thermalization between the fluids (see~\cite{Bertone:2004pz} for a discussion on thermalization and the interactions between DM and OM within the standard cosmological model framework). Comparing the curves, we observe that the $n_D$ increases during neutrino diffusion and neutrino transparent stages when the star is relatively hotter, thus affirming that temperature and neutrino diffusion enhance DM content as observed in Fig~\ref{pf}. In the lower panel, the ratio $n_B/n_0$ increases with the DM chemical potential $\mu_D$, indicating the influence of rising DM energy with density, which can directly affect the physical properties of the star as discussed above.

\begin{figure}
  \includegraphics[scale=1]{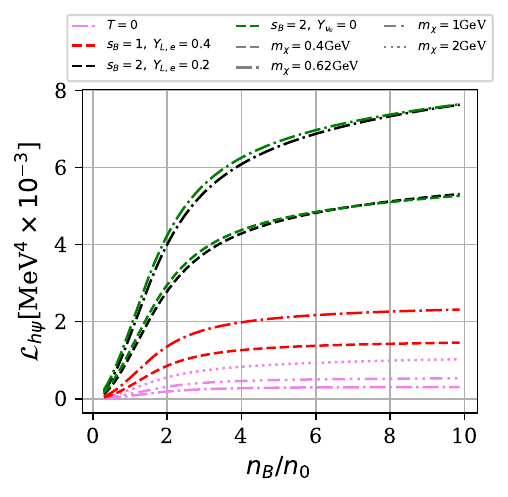}
\caption{A plot of the interaction strength of the Higgs boson $h$ with Standard Model fermions via Yukawa couplings in hot and cold stellar matter. The results are shown for $T=0$ (purple lines) with $m_\chi = 0.62\,\text{GeV}$ (dot-dashed line), $m_\chi = 1\,\text{GeV}$ (dash-double-dotted line) and $m_\chi = 2\,\text{GeV}$ (dotted line), and for hot matter in each case of the pair $\{s_B,Y_{L,e}\}$ equals to $\{1,0.4\}$ (red), $\{2,0.2\}$ (black) and $\{2,0\}$ (green) for $m_\chi = 0.4\,\text{GeV}$ (dashed lines) and $m_\chi = 0.62\,\text{GeV}$ (dot-dashed lines).}
 \label{cpl1}
\end{figure}

\subsubsection{Higgs-OM interaction density}

Figure~\ref{cpl1} illustrates the interaction between the Higgs boson and Standard Model fermions via the Yukawa coupling, as given in Eq.~(\ref{FDMEOS}):
\begin{equation}
    \mathcal{L}_{h\psi} = \sum_{N}\dfrac{fm_N}{v} \bar{\psi}_Nh\psi_N.
\end{equation}
The Higgs boson effectively mediates the interaction between DM and OM through an indirect coupling, where DM interacts with OM via Higgs exchange rather than a new fundamental force. Our results indicate that as temperature increases and neutrinos decouple from the stellar core, the Higgs–OM interaction becomes more pronounced, thereby enhancing the effective DM–OM interaction. Conversely, as the star cools and approaches a cold, catalyzed configuration, this interaction diminishes significantly. This behavior is reflected in the nearly indistinguishable particle fractions shown in Fig.~\ref{pf} (bottom right), as well as the suppressed DM number density seen in the bottom panel of Fig.~\ref{Vden} for cold stars. These effects have been partially addressed in the literature. For instance,~\cite{DeRomeri:2020wng} investigates the thermal production of DM, focusing on a neutralino candidate with $m_\chi \lesssim 1$ GeV, and highlights the temperature dependence of Higgs-mediated interactions. Additional insights into how extreme astrophysical environments influence such interactions are discussed in~\cite{Kouvaris:2014uoa, Arcadi:2019lka}.

\section{Conclusion}\label{c}
We investigated the effects of DM on PNSs using a single-fluid approach, assuming that DM interacts with OM via the Higgs portal and remains in thermal equilibrium due to the non-gravitational interaction between them (see Ref.~\cite{Bertone:2004pz} on thermalization between DM and OM). Our results show that increasing the DM content softens the EoS, leading to more compact stars with lower maximum masses~\cite{Das:2020vng}. DM also absorbs thermal energy from the stellar matter without efficiently re-emitting it, effectively acting as a heat reservoir. This reduces the core temperature and enhances neutrino retention by increasing $x_{\nu_c}$ and neutrino trapping. Discussions on the impact of DM on neutron star cooling and neutrino production can be found in Refs.~\cite{Bhat:2019tnz, Kouvaris:2007ay, Arguelles:2022nbl}. These effects are most significant during deleptonization, when neutrino pressure decreases, making the gravitational influence of DM more pronounced. 

The presence of neutrinos provides additional pressure that helps counteract gravitational collapse, leading to significant changes in particle distribution and the star's structure after all neutrinos have escaped the core (the third stage of stellar evolution). In the cold star configurations, we observe that the star can accrete a larger amount of DM mass while maintaining nearly indistinguishable particle distribution profiles and still satisfying the 2$M_\odot$ observational threshold. This behavior is attributed to reduced thermal pressure, increased central baryon density, and enhanced gravitational binding energy at this stage, all of which contribute to greater compactness and structural stability. Moreover, due to the Higgs-mediated DM–OM interaction, the coupling between DM and OM weakens significantly as the star cools into a cold and catalyzed configuration. This diminished interaction strength likely contributes to the uniformity in the particle composition observed in this phase.

Moreover, our model satisfies key observational constraints, including mass and radius measurements from NICER~\cite{riley2021, Miller:2021qha, riley2019, Miller:2019cac}, as well as the binary NS masses inferred from the gravitational wave signal GW170817~\cite{LIGOScientific:2017vwq, LIGOScientific:2018cki}. When the DM mass is increased to 2 GeV, the results also show an agreement with the HESS data constrained at 90\% CL.  Our results provide insight into how DM can impact NS formation, thermal evolution, and observable properties such as mass and radius, offering potential new constraints on DM mass in supernova remnant evolution. Our findings further shows that to obtain PNSs within the $2\, \rm M_\odot$ maximum mass constraint, a few GeV-scale neutralino mass constraint is required, and in our specific case, we determined that a suitable upper limit of $m_\chi \leq 0.62$ GeV is required from birth to the neutrino-transparent stage to satisfy the $2\, \rm M_\odot$ observable constraint. When we increased the DM mass to $0.62$ GeV, we obtained a maximum mass of 1.99 $M_\odot$ falling slightly under the $2\, \rm M_\odot$ threshold.  A discussion in~\cite{Kamiokande-II:1987idp} places a constraint on the neutralino mass, \( m_\chi \lesssim 1 \) GeV, based on observations from the SN1987A supernova event, which is in good agreement with our results. In the framework of our study, this constraint changes significantly when the star is cooling to a cold NS configuration.

Below, we summarize our main findings and their potential to inform new DM constraints based on observational outcomes:
\begin{itemize}
    \item The alteration of the particle distribution inside the PNSs, particularly in the second and third stages, influenced by the presence of DM, impacts the star's internal density profile (see the value of $n_c$ in Tab.~\ref{ma}), neutrino emissions (see $x_{\nu_e}$), cooling rates (see $T_c$), and also angular momentum. These changes can affect the star's rotational rate, thermal history, response to tidal forces, and gravitational wave signature. Such properties provide valuable observational constraints that could aid in probing the fundamental physics of dense matter and constraints on DM properties.

    \item We find that increasing the DM mass leads to a decrease in the core temperature of the NS, suggesting that DM acts as a heat reservoir, absorbing thermal energy without efficient re-emission. These could alter the cooling curves, deviating them from conventional models and potentially affecting magnetic field evolution. Such effects can be observed through X-ray telescopes (e.g., NICER and future eXTP), indirect neutrino observations, and gravitational wave signals. These deviations, when detected, could provide new constraints on DM properties and its role in NSs.

    \item We found that increasing the DM mass compresses the NS matter, reducing its mass and radius while increasing its compactness due to the stronger gravitational pull. These structural changes could lead to observable effects, such as modifications in gravitational wave signals, X-ray emission profiles, and cooling curves. Comparing these observables with theoretical models can help reveal how DM influences compact astrophysical objects and provide new constraints. 
    
    \item The analysis of the DM–OM coupling properties reveals that DM–OM interactions are significantly enhanced during deleptonization and the neutrino-transparent phase, where the temperature of the stellar matter is higher. Both the $n_D$ and $\mu_D$ exhibit notable increases with rising $s_B$ and decreasing electron $Y_{L,e}$, consistent with a higher temperature regime. Furthermore, the behavior of the Higgs–OM interaction term $\mathcal{L}_{h\psi}$ shows a marked reduction in {interaction density}
    as the star cools into a cold and catalyzed configuration.
\end{itemize}
A comparison between our results and those reported in Ref.~\cite{Issifu:2024htq}, which investigated PNSs admixed with mirror DM using a two-fluid formalism with purely gravitational coupling, reveals important differences in thermal and compositional evolution. While both studies observe similar structural modifications, Ref.~\cite{Issifu:2024htq} found that DM acts as a heat source, raising the core temperature and reducing the isospin asymmetry by enhancing proton production and suppressing neutron production, as seen in the particle fraction profiles. In contrast, the present study, which incorporates non-gravitational DM–OM interactions, shows that DM acts more like a heat sink, leading to a suppression of the temperature distribution. Additionally, we observe an increase in neutron production and a reduction in proton content, resulting in a higher isospin asymmetry. These contrasting results highlight the fundamental role that the nature of DM–OM coupling plays in shaping the thermal and compositional evolution of NSs.

While this study focused on the DDME2 EoS to ensure consistency with astrophysical observations and nuclear saturation properties \cite{PhysRevC.71.024312}, future work will examine the sensitivity of our results to the choice of EoS. In particular, extending the analysis to include softer models such as SFHo \cite{Steiner:2012rk} and intermediate-stiffness models such as TM1 \cite{Sugahara:1993wz} (see \cite{Oertel:2016bki} for a comparison with DDME2) will help assess the robustness of the DM effects discussed here across a broader range of nuclear matter descriptions. Softer EoSs like SFHo are expected to yield smaller maximum masses and reduced stiffness, leading to an earlier onset of instability under DM admixture and potentially lowering the derived $m_\chi$ limit. Stiffer EoSs beyond DDME2 (e.g., NL3 \cite{Lalazissis:1996rd}) should support larger maximum masses and delay instability, likely allowing greater DM masses. A systematic scan over multiple EoS parameterizations, therefore, remains an important extension of this work.

 While this study focused on the DDME2 EoS to ensure consistency with astrophysical observations and nuclear saturation properties \cite{PhysRevC.71.024312}, future work will examine the sensitivity of our results to the choice of EoS. In particular, extending the analysis to include softer models such as SFHo \cite{Steiner:2012rk} and intermediate-stiffness models such as TM1 \cite{Sugahara:1993wz} (see \cite{Oertel:2016bki} for a comparison with DDME2) will help assess the robustness of the DM effects discussed here across a broader range of nuclear matter descriptions. Softer EoSs like SFHo are expected to yield smaller maximum masses and reduced stiffness, leading to an earlier onset of instability under DM admixture and potentially lowering the derived $m_\chi$ limit. Stiffer EoSs beyond DDME2 (e.g., NL3 \cite{Lalazissis:1996rd}) should support larger maximum masses and delay instability, likely allowing greater DM masses. A systematic scan over multiple EoS parameterizations, therefore, remains an important extension of this work.

\begin{acknowledgments}

A.I. acknowledges financial support from the São Paulo State Research Foundation (FAPESP), Grant No. 2023/09545-1. 
This work is part of the project INCT-FNA (Proc. No. 464898/2014-5) and is also supported by the National Council for Scientific and Technological Development (CNPq) under Grants Nos. 303490/2021-7 (D.P.M.) and 306834/2022-7 (T.F.) T. F. also thanks the financial support from  Improvement of Higher Education Personnel CAPES (Finance Code 001) and FAPESP Thematic Grants (2023/13749-1 and 2024/17816-8). 

\end{acknowledgments}

\bibliography{refs}

\end{document}